\documentclass[onecolumn,showpacs]{revtex4}

\usepackage{graphicx}% Include figure files
\usepackage{dcolumn}% Align table columns on decimal point
\usepackage{bm}% bold math

\input epsf

\begin{document}

\title{\Large Brans-Dicke  Cosmology  in  an  anisotropic  model \\
when  Velocity  of  Light  Varies}

\author{\bf Subenoy Chakraborty}
\email{subenoyc@yahoo.co.in}
\author{\bf N.C. Chakraborty}
\author{\bf Ujjal Debnath}
\email{ujjaldebnath@yahoo.com}

\affiliation{Department of Mathematics, Jadavpur University,
Calcutta-32, India.}

\date{\today}

\begin{abstract}
In  this  paper, we  have  studied  Brans-Dicke  Cosmology  in
anisotropic  Kantowski-Sachs  space-time  model; considering
variation  of  the  velocity  of  light. We  have  addressed the
flatness  problem  considering  both  the  cases  namely, (i)
when  the  Brans-Dicke  scale  field $\phi$ is  constant (ii)
when $\phi$ varies, specially  for  radiation  dominated  era
perturbatively and  non-perturbatively  and  asymptotic
behaviour  have  been studied.
\end{abstract}

\maketitle

\section{\normalsize\bf{Introduction}}
 In  recent  years  theories  with  varying  speed  of  light  (VSL)  has  been  attracted
 considerable  attention  [1-10]  due  to  its  ability  to  solve  the  so-called
 cosmological  puzzles -  the  horizon, flatness  and  Lambda  problems  of  big-bang
 cosmology. All  previous  attempts  to  overcome  these  difficulties  invoke  the  basic
 inflationary  form, where  the  observable  universe  experiences  a  period  of
 `superluminal'  expansion. Here,  one  has  to  modify  the  matter  content  of  the
 universe  such  that  ordinary  Einstein  gravity  becomes  repulsive  and  halts  the
 exponential  expansion. One  can  however resolve to theories  with  varying  speed  of
 light  as  the  alternative  method  to  solve  the  above  mentioned  cosmological
 puzzles. Here, instead  of  changing  the  matter  content  of  the  universe, one
 has  to  change  the  speed  of  light  in  the  early  universe. The  universe  is
 assumed  to  be  radiation  dominated  in the  early  stage  and  the  matter  content
 is assumed to be the  same  as  in  the  standard  big-bang  (SBB)  model. So  the  geometry  and
 expansion  factor  of  the  universe  go  in  accordance  with  the  SBB  model.
 Here for a  free  falling  observer  associated  with  the  cosmic  expansion,  the
 local  speed  of  light  will decelerate  from  a  very  large  value  to  its  current
 value.\\

 So  far  some  works  have  been  done  with  VSL  by  Magueijo  and  co-workers [1-5] and
 others [6-10]  specially  in  isotropic  cosmological models (which also include BD-theory).
 Very  recently, Magueijo  has  investigated  [11], the possibility  of  black  holes formation
 by  studying  spherically  symmetric  solutions with  VSL. In  this  paper,
 we  will  investigate  in  details  the  anisotropic  Kantowski-Sachs (KS) model  with
 VSL  theory. The  paper  is  organized  as  follows:  section II deals  with the basic
 equations  in  BD-theory  for  KS  model  with  varying  speed  of  light. Perturbative
 solution  to  the  flatness  problem  will be discussed  in  section III, while
 non-perturbative  study  of  this  problem  will be  discussed  in  section IV  for
 radiation  era  only  and  the  exact  solution  to  the  flatness  problem  is
 presented  in  section V. In  section VI, we  will  study  solutions  to  the
 quasi-flatness  problem  and  examine its asymptotic  behaviour. The
 lambda  and  quasi-lambda  problems  and their asymptotic  behaviours  will
 be investigated  in  section VII. The  paper  concludes  with  a  short  discussion
 in section VIII.\\

\section{\normalsize\bf{The  basic  equations }}
For  anisotropic  KS  model  with  metric  ansatz

$$
ds^{2}=-c^{2}dt^{2}+a^{2}(t)dr^{2}+b^{2}(t)d\Omega_{2}^{2}~,
$$

the  BD  field  equations  with  varying  speed  of  light  are

\begin{equation}
\frac{\ddot{a}}{a}+2\frac{\ddot{b}}{b}=-\frac{8\pi}{(3+2\omega)\phi}\left[
(2+\omega)\rho+3(1+\omega)\frac{p}{c^{2}}\right]-\omega\left(\frac{\dot{\phi}}{\phi}
\right)^{2}-\frac{\ddot{\phi}}{\phi}
\end{equation}

\begin{equation}
\left(\frac{\dot{b}}{b}
\right)^{2}+2\frac{\dot{a}}{a}\frac{\dot{b}}{b}=\frac{8\pi\rho}{\phi}-\frac{c^{2}}{b^{2}}-\left(\frac{\dot{a}}{a}+2\frac{\dot{b}}{b}
\right)\frac{\dot{\phi}}{\phi}+\frac{\omega}{2}\left(\frac{\dot{\phi}}{\phi}
\right)^{2}
\end{equation}

and  the  wave  equation  is

\begin{equation}
\ddot{\phi}+\left(\frac{\dot{a}}{a}+2\frac{\dot{b}}{b}
\right)\dot{\phi}=\frac{8\pi}{3+2\omega}\left(\rho-\frac{3p}{c^{2}}
\right)
\end{equation}

Here  the  velocity  of  light  $c$  is  an  arbitrary  function
of  time , $\omega$ is  the  BD  coupling  parameter  and  the BD
scalar  field  is $\phi=\frac{1}{G}$. From  the  above  field
equations  we have  the  ``non-conservation''  equation

\begin{equation}
\dot{\rho}+\left(\frac{\dot{a}}{a}+2\frac{\dot{b}}{b}
\right)\left(\rho+\frac{p}{c^{2}} \right)=\frac{c\dot{c}}{4\pi
b^{2}}\phi
\end{equation}

Now  $p=\frac{1}{3}\rho c^{2}$ is  the  equation  of  state  in
the radiation  era for  which  the  general  solution  $\phi$ is
(from equation (3) )

\begin{equation}
\phi=\phi_{0}+\alpha\int{\frac{dt}{a b^{2}}}.
\end{equation}

The  above  field  equations  in  BD-theory  have  been
formulated  in  Jordan  frame. To  switch  over  to  Einstein
frame  we  shall  have  to  make  the  following  transformations

\begin{eqnarray*}
d\hat{t}=\sqrt{G\phi}~dt,~~\hat{a}=\sqrt{G\phi}~a,~~\hat{b}=\sqrt{G\phi}~b,
~~\sigma=\left(\omega+\frac{3}{2}\right)^{1/2}ln(G\phi),~~\hat{\rho}=(G\phi)^{-2}\rho,
\end{eqnarray*}
\vspace{-8mm}

\begin{equation}
~~\hat{p}=(G\phi)^{-2}p~,\hspace{-3in}
\end{equation}

and  the  above  field  equations  become ($c$ is  treated  as
constant)

\begin{equation}
\frac{\hat{a}''}{\hat{a}}+2\frac{\hat{b}''}{\hat{b}}=-4\pi
G\left(\hat{\rho}+\frac{3\hat{p}}{c^{2}}\right)-\sigma'^{2}~,
\end{equation}

\begin{equation}
\left(\frac{\hat{b}'}{\hat{b}}\right)^{2}+2\frac{\hat{a}'}{\hat{a}}\frac{\hat{b}'}{\hat{b}}
+\frac{c^{2}}{\hat{b}^{2}}=8\pi
G\hat{\rho}+\frac{\hat{\sigma}'^{2}}{2}~,
\end{equation}

and

\begin{equation}
\sigma''+\left(\frac{\hat{a}'}{\hat{a}}+2\frac{\hat{b}'}{\hat{b}}\right)\sigma'=\frac{8\pi
G}{\sqrt{6+4\omega}}\left(\hat{\rho}-\frac{3\hat{p}}{c^{2}}\right)
\end{equation}

with  the prime `` $'$ '' and `` . '' represent the
differentiation with respect to $\hat{t}$ and $t$ respectively.\\

One  can  interpret  these  field  equations  as  standard  KS
equations  with  constant $G$  and  a  scalar  field  $\sigma$  is
added to  the  normal  matter . The  scalar  field  behaves like
a ``stiff''  perfect  fluid  with  equation  of  state

\begin{equation}
\hat{p}_{\sigma}=\hat{\rho}_{\sigma}=\frac{\sigma'^{2}}{16\pi G}~.
\end{equation}

If  the  velocity  of  light  is  constant , then  in  Einstein
frame  total  stress-energy  tensor  is  conserved  but  there
is  an  exchange  of  energy  between  the  scalar  field  and
normal  matter  according  to  the  following  equation

\begin{equation}
\hat{\rho}'+\left(\frac{\hat{a}'}{\hat{a}}+2\frac{\hat{b}'}{\hat{b}}\right)
\left(\hat{\rho}+\frac{\hat{p}}{c^{2}}\right)=-\left[\hat{\rho}'_{\sigma}+
\left(\frac{\hat{a}'}{\hat{a}}+2\frac{\hat{b}'}{\hat{b}}\right)\left(
\hat{\rho}_{\sigma}+\frac{\hat{p}_{\sigma}}{c^{2}}\right)\right]=
-\frac{\sigma'}{\sqrt{6+4\omega}}\left(\hat{\rho}-\frac{3\hat{p}}{c^{2}}\right)
\end{equation}

On  the  other  hand , if  the  velocity  of  light  varies then
we  have two separate  ``non-conservation'' equations

\begin{equation}
\hat{\rho}'+\left(\frac{\hat{a}'}{\hat{a}}+2\frac{\hat{b}'}{\hat{b}}\right)
\left(\hat{\rho}+\frac{\hat{p}}{c^{2}}\right)=-\frac{\sigma'}{\sqrt{6+4\omega}}
\left(\hat{\rho}-\frac{3\hat{p}}{c^{2}}\right)+\frac{c c'}{4\pi
G\hat{b}^{2}}~,
\end{equation}

and

\begin{equation}
\hat{\rho}'_{\sigma}+
\left(\frac{\hat{a}'}{\hat{a}}+2\frac{\hat{b}'}{\hat{b}}\right)\left(
\hat{\rho}_{\sigma}+\frac{\hat{p}_{\sigma}}{c^{2}}\right)=\frac{\sigma'}
{\sqrt{6+4\omega}}\left(\hat{\rho}-\frac{3\hat{p}}{c^{2}}\right)
\end{equation}

Thus  standard  KS  model  field  equations  with  constant  $G$
for  VSL  are  also  the  field  equations  for  BD  theory  in
Einstein  frame  but  here  one  must  add  to  normal  matter
the  scalar  field  energy  and  pressure . Therefore , the
total  energy  density  and  pressure  are  given  by
$\hat{\rho}_{t}=\hat{\rho}+\hat{\rho}_{\sigma}$ and
$\hat{p}_{t}=\hat{p}+\hat{p}_{\sigma}$ respectively .

\section{Perturbative  solutions  to  the  flatness  problem}

For  a  possible  solution  to  the  flatness  problem  in  VSL
theory , let us  first  study  solutions  when  there  are  small
deviations  from  flatness . The  critical  energy  density
$\rho_{c}$ in a  BD  universe  is  given  by  the  equation

\begin{equation}
\left(\frac{\dot{b}}{b}
\right)^{2}+2\frac{\dot{a}}{a}\frac{\dot{b}}{b}=\frac{8\pi\rho_{c}}{\phi}-\left(\frac{\dot{a}}{a}+2\frac{\dot{b}}{b}
\right)\frac{\dot{\phi}}{\phi}+\frac{\omega}{2}\left(\frac{\dot{\phi}}{\phi}
\right)^{2}
\end{equation}

In  Einstein  frame  the  critical  energy  density  for  normal
matter  is

\begin{equation}
\hat{\rho}_{c}=\frac{1}{8\pi G
}\left[\left(\frac{\hat{b}}{\hat{b}}\right)^{2}+2\frac{\hat{a}'}{\hat{a}}\frac{\hat{b}'}{\hat{b}}
-\frac{\hat{\sigma}'^{2}}{2}\right]=\frac{\rho_{c}}{G^{2}\phi^{2}}
\end{equation}

Hence,if we define  the  total  critical  energy  as

$$
\hat{p}_{\alpha}=\hat{\rho}_{c}+\hat{\rho}_{\sigma}~,
$$

then  from  (15)  we  have

\begin{equation}
\hat{\rho}_{\alpha}=\frac{1}{8\pi
G}\left[\left(\frac{\hat{b}}{\hat{b}}
\right)^{2}+2\frac{\hat{a}'}{\hat{a}}\frac{\hat{b}'}{\hat{b}}\right]
\end{equation}

Let  us  define  a  relative  flatness  parameter  as

\begin{equation}
\varepsilon_{t}=\frac{\hat{\rho}_{t}-\hat{\rho}_{\alpha}}{\hat{\rho}_{\alpha}}
\end{equation}

whose  evolution  equation  is given by

\begin{equation}
\hat{\varepsilon}'_{t}=(1+\hat{\varepsilon}_{t})\hat{\varepsilon}_{t}\left(
\gamma\frac{\hat{a}'}{\hat{a}}+2(\gamma-1)\frac{\hat{b}'}{\hat{b}}\right)
+2\frac{c'}{c}\hat{\varepsilon}_{t}
\end{equation}

Here $\gamma-1=p/\rho c^{2}=\hat{p}/\hat{\rho}c^{2}$ , is  the
equation of state, with $\gamma$ constant. Since

\begin{equation}
\hat{\varepsilon}'_{t}=\frac{\varepsilon}{1+(2\omega+3)\dot{\phi}^{2}/32\pi\phi\rho_{c}}
\end{equation}

in  Jordan  frame,  the  deviation  from  flatness  is  given  by

\begin{equation}
\delta=\frac{\rho-\rho_{c}}{\rho_{c}+(2\omega+3)\dot{\phi}^{2}/32\pi\phi\rho_{c}}<\varepsilon
\end{equation}

where $\varepsilon=(\rho-\rho_{c})/\rho_{c}$  measures  natural
deviations from flatness. Now $\delta$, the  adaptation  parameter
satisfies  the differential  equation

\begin{equation}
\dot{\delta}=\delta(1+\delta)\left(\gamma\frac{\dot{a}}{a}+2(\gamma-1)\frac{\dot{b}}{b}
+\frac{1}{2}(3\gamma-2)\frac{\dot{\phi}}{\phi}\right)+2\frac{\dot{c}}{c}\delta
\end{equation}

If $\delta$ is  assumed  to  be  very  small  compare  to  unity
i.e.,$\delta\ll 1$, then  neglecting  the  square  term in (21),
one can integrate the above to  get

\begin{equation}
\delta=\delta_{0}a^{\gamma}b^{2(\gamma-1)}\phi^{(3\gamma-2)/2}c^{2}
\end{equation}

where $\delta_{0}$ is the integration  constant. Hence we have

\begin{equation}
\frac{1}{\varepsilon}=\frac{\varepsilon_{0}}{a^{\gamma}b^{2(\gamma-1)}
\phi^{(3\gamma-2)/2}c^{2}}-\frac{(2\omega+3)\dot{\phi}^{2}}{32\pi\rho_{c}\varepsilon}
\end{equation}

In  BD  theory $G\propto\phi^{-1}$ , so  to  solve  the flatness
problem $\phi$ should decrease  in  the  early  universe and the
first term on  the right  hand  side  must  dominate  over
the second one.\\

In  particular, in  radiation  era  ($\gamma=4/3$)  the  above
equation (23)  simplifies  to

\begin{equation}
\frac{1}{\varepsilon}=\frac{\varepsilon_{0}}{a^{4/3}b^{2/3}\phi
c^{2}}-\frac{(2\omega+3)\alpha^{2}}{32\pi\phi
a^{2}b^{4}\rho_{c}\varepsilon}
\end{equation}

where  we  have  used  equation  (5)  to  obtain  $\phi$ . If
$\alpha$ is small  and
$c=c_{0}\left[a^{n}b^{2(n-1)}\right]^{n/(3n-2)}$ then [12,3] it is
still possible to  solve the  flatness  problem for
$n<-1-\sqrt{7/3}$ . Furthermore, since $\rho_{c}\propto
1/(ab^{2})^{4/3}$ at late times, so the second term is always very
small compare to the  first and hence may be neglected.\\

\section{Non-perturbative  solutions  with  $\phi$ = constant:~ Matter
and Radiation  dominated  era}

When $\phi=\phi_{0}$ (constant) we have $\delta=\varepsilon$ in
(20) and the differential equation for $\delta$, (21) can be
written as

\begin{equation}
\dot{\varepsilon}=\varepsilon(1+\varepsilon)\left(\gamma\frac{\dot{a}}{a}
+2(\gamma-1)\frac{\dot{b}}{b}\right)+2\frac{\dot{c}}{c}\varepsilon
\end{equation}

Now, if  we  assume a  power-law  form  in  the scale  factors
of  the  velocity  of  light  then  the  above equation  becomes

\begin{equation}
\dot{\varepsilon}=\varepsilon(1+\varepsilon)\left(\gamma\frac{\dot{a}}{a}
+2(\gamma-1)\frac{\dot{b}}{b}\right)+2\varepsilon\left(\frac{n}{3n-2}\right)
\left(n\frac{\dot{a}}{a}+2(n-1)\frac{\dot{b}}{b}\right)
\end{equation}

In  order  to  get  an  exact  analytic  solution  to  the above
equation, we  assume  $\varepsilon\ll 1$ , so  we can neglect
$\varepsilon^{2}$-term. We then have  the  solution

\begin{equation}
\varepsilon=\varepsilon_{0}a^{\gamma+2n^{2}/(3n-2)}b^{2(\gamma-1)+4n(n-1)/(3n-2)}
\end{equation}

with $\varepsilon_{0}$ as the  integration  constant.\\

Furthermore, without  assuming  any  restriction  on  the
parameter $\varepsilon$, if  we  assume  the  two  arbitrary
constants $n$ and  $\gamma$ to  be  equal  then  also  we have an
exact  analytic solution  for $\varepsilon$,
\begin{equation}
\varepsilon=-\frac{1}{\frac{3\gamma-2}{5\gamma-2}+
A\left[a^{\gamma}b^{2(\gamma-1)}\right]^{-(5\gamma-2)/(3\gamma-2)}}
\end{equation}

where $A$  is  an  integration  constant.\\

\begin{figure}
\includegraphics{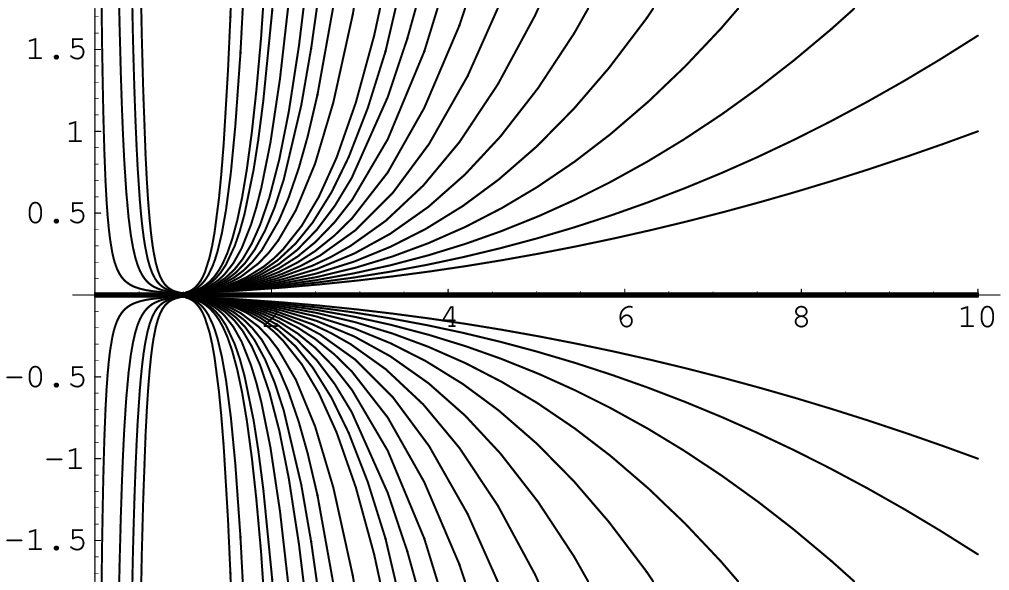}
\caption{Here the flatness parameter $\varepsilon$ in (27) has
been plotted against $t$ for the radiation era ($\gamma=4/3$)
choosing different values for the constant $n$. The scale factors
here are assumed to be in simple expanding form.\\\\}
\includegraphics{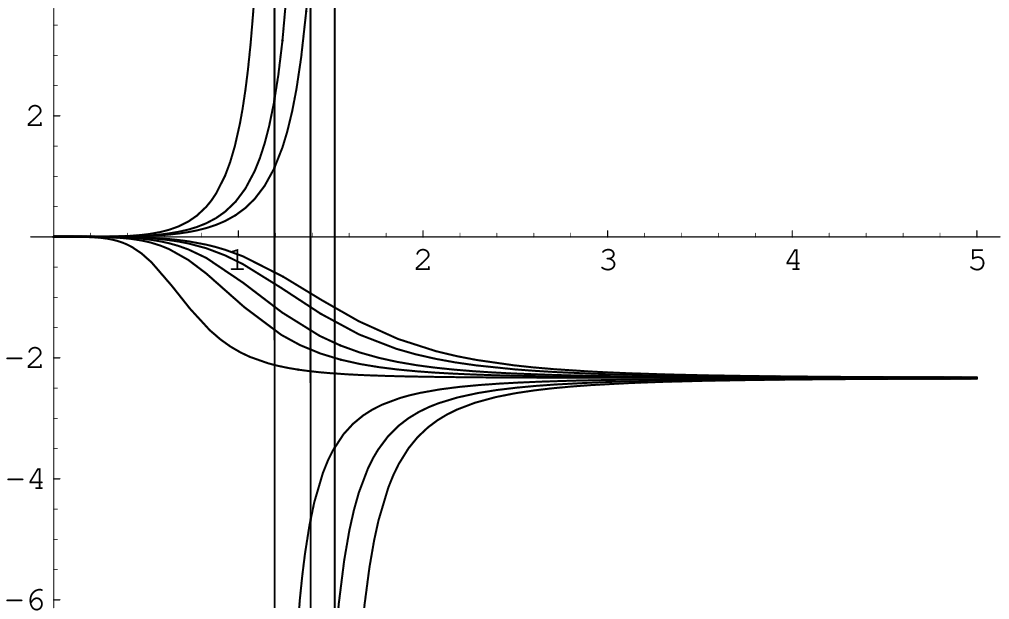}
\caption{This figure shows the variation of the flatness
parameter from (28), where there is no restriction on
$\varepsilon$. This figure is also drawn for the radiation era
and assuming as before the simple expanding form for the scale
factors. The variation of $\varepsilon$ for different choice of
the arbitrary constant $A$ is shown in the figure.}
\end{figure}

\section{Exact  solutions  to  the  flatness  problem:~ Radiation
dominated  era}

To  obtain  the  exact  solutions  to  the  flatness  problem we
assume  the  variation  of  light  mathematically  by  the
relation [3,12]

\begin{equation}
c=c_{0}\left[\left\{a^{n}b^{2(n-1)}\right\}^{1/(3n-2)}\sqrt{\phi G
}\right]^{n}
\end{equation}

However, in  the  Einstein  frame  the  above  relation reduces
to  $c=c_{0}\left[\hat{a}^{n}\hat{b}^{2(n-1)}\right]^{1/(3n-2)}$
and equation (12) becomes

\begin{equation}
\hat{\rho}'+\left(\frac{\hat{a}'}{\hat{a}}+2\frac{\hat{b}'}{\hat{b}}\right)
\left(\hat{\rho}+\frac{\hat{p}}{c^{2}}\right)=\frac{cc'}{4\pi
G\hat{b}^{2}}
\end{equation}

The  equation  of  state $\hat{p}=(\gamma-1)\hat{\rho}c^{2}$
simplifies the above equation further to

\begin{equation}
\hat{\rho}'+\gamma\hat{\rho}\left(\frac{\hat{a}'}{\hat{a}}+2\frac{\hat{b}'}{\hat{b}}\right)
=\frac{cc'}{4\pi G\hat{b}^{2}}
\end{equation}

An  exact  integral  can  be  obtained (assuming $n=\gamma$) as

\begin{equation}
\hat{\rho}'(\hat{a}\hat{b}^{2})^{\gamma}=B+\frac{\gamma
c_{0}^{2}}{4\pi
G(5\gamma-2)}\left[\hat{a}^{\gamma}\hat{b}^{2(\gamma-1)}\right]^{(5\gamma-2)/(3\gamma-2)}
\end{equation}

Hence  for  the  radiation  dominated  era , we  have  in
Jordan  frame

\begin{equation}
\rho(ab^{2})^{4/3}=B+\frac{c_{0}^{2}}{14\pi G}(a^{2}b)^{14/9}(\phi
G)^{7/3}
\end{equation}

where  we  have  used  $\hat{\rho}(\hat{a}\hat{b}^{2})^{4/3}=\rho(ab^{2})^{4/3}$ .\\

Moreover, in the Jordan  frame  we  have  equation  (29)  for  the
velocity of  light  and  we  can  write

\begin{equation}
\frac{\dot{c}}{c}=\frac{n}{3n-2}\left[n\frac{\dot{a}}{a}+2(n-1)\frac{\dot{b}}{b}\right]+
\frac{n}{2}\frac{\dot{\phi}}{\phi}
\end{equation}

Thus, if  $\phi$ is a decreasing function  then
$|\dot{c}/c|<|[n\dot{a}/a+2(n-1)\dot{b}/b]/(3n-2)|$ and we can
therefore conclude  that a varying speed of light in  the  early
universe (with weaker gravity) helps to solve  the flatness
problem.\\

Alternatively, for  general  equation  of  state
$p=(\gamma-1)\rho c^{2}$ , if  we make  the  transformation

\begin{equation}
x=\phi a^{\gamma}b^{2(\gamma-1)}
\end{equation}

and  assume  the  variation  of  the  velocity  of  light  as

$$
c=c_{0}x^{n/2}
$$

then  the  equation  of  continuity  (4)  becomes

\begin{equation}
\dot{\rho}+\gamma\rho\left(\frac{\dot{a}}{a}+2\frac{\dot{b}}{b}\right)=
\frac{nc_{0}^{2}\phi x^{n-1}\dot{x}}{8\pi b^{2}}
\end{equation}

Integrating this  equation  gives

\begin{equation}
\rho
a^{\gamma}b^{2\gamma}=B+\frac{nc_{0}^{2}\phi^{n+1}\left[a^{\gamma}
b^{2(\gamma-1)}\right]^{n+1}}{8\pi(n+1)}
\end{equation}

(provided $n\ne -1$) which gives  the  evolution  for  general
$\gamma$.

\section{Solutions  to  the  quasi-flatness  problem}

\subsection{When BD scalar  field  $\phi$ is  constant}

In  this  section  we  shall  investigate  whether  it  is
natural  to  have  evolution  which  asymptotes  to  a  state of
expansion  with  a  non-critical  density . Let  us  start with
the  continuity  equation  for  constant  $\phi$  and  assume as
before  the  equation  of  state

$$
\frac{p}{\rho c^{2}}=\gamma-1~,
$$

and  the  velocity  of  light to be

$$
c=c_{0}\left[a^{n}b^{2(n-1)}\right]^{n/(3n-2)}
$$

then  the  integral  gives

\begin{equation}
\rho(ab^{2})^{\gamma}=B+\frac{\gamma
c_{0}^{2}\phi}{4\pi(5\gamma-2)}\left[a^{\gamma}b^{2(\gamma-1)}\right]^{(5\gamma-2)/(3\gamma-2)}
\end{equation}

(It  is  to  be  noted  that  in  order  to  obtain  the
integral  one  has  to  assume $n=\gamma$).\\

Now  substituting  this  value  of  $\rho$ in  equation (2) we
have

\begin{eqnarray*}
\left(\frac{\dot{b}}{b}\right)^{2}+2\frac{\dot{a}}{a}\frac{\dot{b}}{b}
=\frac{8\pi}{\phi}-\frac{B}{(ab^{2})^{\gamma}} +\frac{\gamma
c_{0}^{2}\phi}{4\pi(5\gamma-2)}+\frac{2\left[a^{\gamma}b^{2(\gamma-1)}
\right]^{(5\gamma-2)/(3\gamma-2)}}{(ab^{2})^{\gamma}}
\end{eqnarray*}
\vspace{-5mm}

\begin{equation}
-\frac{1}{b^{2}}c_{0}^{2}\left[a^{\gamma}b^{2(\gamma-1)}
\right]^{2\gamma/(3\gamma-2)}
\end{equation}

In  particular, for  radiation  era  the  above  equation
simplifies  to

\begin{equation}
\left(\frac{\dot{b}}{b}\right)^{2}+2\frac{\dot{a}}{a}\frac{\dot{b}}{b}
=\frac{8\pi B}{\phi
(ab^{2})^{4/3}}-\frac{3}{7}c_{0}^{2}\left(a^{8}b^{-5}\right)^{2/9}
\end{equation}

If $\Omega$ is  the  density  parameter, then  it  can be  defined
as

\begin{equation}
\frac{\Omega}{\Omega-1}=\frac{8\pi G\rho}{c^{2}/b^{2}}
\end{equation}

We  note  that  the  ratio  between  the  two  terms  on  the
right  hand  side  is  almost  a  constant  for  a  quasi-flat
open  universe. However, for $\gamma=4/3$, we  get

\begin{equation}
\frac{\Omega}{\Omega-1}=\frac{4}{7}+\frac{8\pi G
B}{c_{0}^{2}(a^{2}b)^{14/9}}
\end{equation}

For  expanding  universe  $a, b$  grows  with  time. So
asymptotically  the  second  term  on  the  R.H.S  will  be
negligible  and  we  have

$$
\Omega=-\frac{4}{3}
$$

which  leads  to  an  open  universe  with  finite  $\Omega$
value today.

\subsection{When  BD  scalar  field  varies}

For  the  radiation era ($\gamma=4/3$), let  us  define

\begin{equation}
y=\phi (ab^{2})^{2/3},~~~z=\phi(a^{2}b)^{2/3},~~~c=c_{0}z^{m}
\end{equation}

So  from  the  field  equation  (2)  we  get  ( using  equation
(5) )

\begin{equation}
2\frac{y'}{y}\frac{z'}{z}-\frac{z'^{2}}{z^{2}}=\frac{32\pi\rho}{3y}(ab^{2})^{4/3}-
\frac{4c^{2}}{3}\frac{z}{y}+\left(1+\frac{2\omega}{3}\right)\frac{\alpha^{2}}{y^{2}}
\end{equation}

Now, from  equation  (38) the expression of $\rho$ takes  the form

\begin{equation}
\rho(ab^{2})^{4/3}=\frac{Bmc_{0}^{2}z^{2m+1}}{4\pi(2m+1)}
\end{equation}

In addition  the  critical  density $\rho_{c}$ is  obtained  from
the differential  equation

\begin{equation}
2\frac{y'}{y}\frac{z'}{z}-\frac{z'^{2}}{z^{2}}=\frac{32\pi\rho_{c}}{3y}(ab^{2})^{4/3}
+\left(1+\frac{2\omega}{3}\right)\frac{\alpha^{2}}{y^{2}}
\end{equation}

Hence  the  expression  for  the  density  parameter  is

$$
\Omega=\frac{\rho}{\rho_{c}}=\frac{2\frac{y'}{y}\frac{z'}{z}+
\frac{4c^{2}}{3}\frac{z}{y}-\left(1+\frac{2\omega}{3}\right)\frac{\alpha^{2}}{y^{2}}}
{2\frac{y'}{y}\frac{z'}{z}-\frac{z'^{2}}{z^{2}}-\left(1+\frac{2\omega}{3}\right)\frac{\alpha^{2}}{y^{2}}}
$$

or  equivalently,

\begin{equation}
\frac{\Omega}{\Omega-1}=\frac{\rho}{\rho-\rho_{c}}=\frac{8\pi
B}{c_{0}^{2}z^{2m+1}}+\frac{2m}{2m+1}
\end{equation}

Thus, if $2m+1<0$  then  as $t\rightarrow\infty,
z\rightarrow\infty$ so  we  have $\Omega\rightarrow 1$. But  if
$2m+1>0$ then $t\rightarrow\infty, \Omega\rightarrow -2m$. So
asymptotically, we have  a quasi-flat open universe.

\subsection{General  asymptotic  behaviour}
In  this  section, we  consider  general  asymptotic  behaviour
when $2m+1>0$ i.e., for  quasi-flat  open  universe. The equation
(44) can be approximated  by

$$
2y'z'-\frac{y z'^{2}}{z}\simeq \Gamma~ z^{2(m+1)}
$$

which  has  a  first  integral

$$
\frac{y}{\sqrt{z}}=\int\frac{\Gamma~
z^{2(m+1)}}{2z'\sqrt{z}}~d\tau
$$

Thus, if we assume $z\sim \tau^{2/\Omega_{\infty}}$, then  $y\sim
\tau^{2/\Omega_{\infty}}$ and

\begin{equation}
\phi\sim\phi_{0}~exp\left(\frac{\tau^{1-2/\Omega_{\infty}}}{1-
\tau^{2/\Omega_{\infty}}}\right)
\end{equation}

For the case of $\Omega_{\infty}<1$, we have
$\phi\rightarrow\phi_{0},~ a(\tau)\sim \tau^{1/\Omega_{\infty}}$
and $b(\tau)\sim \tau^{1/\Omega_{\infty}}$ in the limit as
$\tau\rightarrow\infty$, or equivalently,

$$
\phi\rightarrow\phi_{0},~~~~a\sim
\tau^{1/(\Omega_{\infty}+1)},~~~~b\sim
\tau^{1/(\Omega_{\infty}+1)},~~~~~ as~ t\rightarrow\infty.
$$

In addition, when $\Omega_{\infty}=1$, we get $a\sim t^{1/2},~
b\sim t^{1/2}$, as expected for flat radiation  asymptote .

\section{The  Lambda  and  the  quasi-lambda  problems}
In  this  section, we  shall  examine  the  effect  of variation
of  speed  of  light  when  a  cosmological  term  is introduced
in  the  BD  field  equations. In  fact, the incorporation  of  a
cosmological  term  is  equivalent  to introduction of  a  vacuum
stress  which obeys  an  equation  of  state

$$
\rho_{\Lambda}=-\frac{p_{\Lambda}}{c^{2}}
$$

with

$$
\rho_{\Lambda}=\frac{\Lambda c^{2}}{8\pi G}
$$

Thus  equation  of  continuity  can  be  generalized  to

$$
(\dot{\rho}+\dot{\rho}_{\Lambda})+\left(\frac{\dot{a}}{a}+2\frac{\dot{b}}{b}\right)
=\frac{c\dot{c}}{4\pi G ^{2}b}
$$

Also  the  field  equation  (2)  can  now  be  written  as

\begin{equation}
\frac{\dot{b}^{2}}{b^{2}}+2\frac{\dot{a}}{a}\frac{\dot{b}}{b}=8\pi
G\rho-\frac{c^{2}}{b^{2}}+\Lambda c^{2}
\end{equation}

Now, using  the  above  expression  for  vacuum  stress-energy
and using $\gamma-1=p/\rho c^{2}$ as  the  equation  of  state
for the matter  and assuming

$$
c=c_{0}\left[a^{n}b^{2(n-1)}\right]^{n/(3n-2)},
$$

the  above  equation  of  continuity  can  be  integrated
(assuming $n=\gamma$) to  give

\begin{equation}
\rho=\frac{B}{(ab^{2})^{\gamma}}+\frac{\gamma c_{0}^{2}}{8\pi
G(5\gamma-2)}\frac{\left[a^{\gamma}b^{2(\gamma-1)}\right]^{\frac{5\gamma-2}{3\gamma-2}}}
{(ab^{2})^{\gamma}}-\frac{\Lambda c_{0}^{2}(5\gamma-2)}{8\pi
G}\frac{1}{ab^{2}}\int(ab^{2})^{\gamma}d[a^{\gamma}b^{2(\gamma-1)}]^{\frac{2\gamma}{3\gamma-2}}
\end{equation}
\\
On substituting  this  value of $\rho$ in  the  above  field
equation (49), we  have

\begin{eqnarray*}
\frac{\dot{b}^{2}}{b^{2}}+2\frac{\dot{a}}{a}\frac{\dot{b}}{b}=\frac{8\pi
G B}{(ab^{2})^{\gamma}}+\frac{2\gamma
c_{0}^{2}}{(5\gamma-2)}\frac{\left[a^{\gamma}b^{2(\gamma-1)}\right]^{(5\gamma-2)/(3\gamma-2)}}
{(ab^{2})^{\gamma}}+\left(\Lambda-\frac{1}{b^{2}}\right)c_{0}^{2}
\left[a^{\gamma}b^{2(\gamma-1)}\right]^{2\gamma/(3\gamma-2)}
\end{eqnarray*}
\vspace{-5mm}

\begin{equation}
-\Lambda
c_{0}^{2}(5\gamma-2)\frac{1}{(ab^{2})^{\gamma}}\int(ab^{2})^{\gamma}d[a^{\gamma}
b^{2(\gamma-1)}]^{2\gamma/(3\gamma-2)}\hspace{-1.5in}
\end{equation}

If  we  define  the  generalized  density  parameter as

\begin{equation}
\Omega_{1}=\Omega_{m}+\Omega_{\Lambda}=\frac{8\pi
G(\rho+\rho_{\Lambda})b^{2}}{c^{2}}
\end{equation}

then  using  the  density  parameter $\Omega$ from  equation
(41) we have

$$
\Omega_{1}=\frac{\Omega}{\Omega-1}=\frac{8\pi
G(\rho+\rho_{\Lambda})b^{2}}{c^{2}}
$$

If  we  now  substitute  the  value  of  $\rho$ from  equation
(50) we obtain

\begin{eqnarray*}
\frac{\Omega}{\Omega-1}=\frac{2\gamma}{5\gamma-2}\left[a^{\gamma}b^{2(\gamma-1)}
\right]^{(4\gamma-3)/(3\gamma-2)}+\frac{8\pi G
B}{c_{0}^{2}}\left[a^{\gamma}b^{2(\gamma-1)}
\right]^{-(4\gamma-2)/(3\gamma-2)}+\Lambda b^{2}
\end{eqnarray*}
\vspace{-5mm}

\begin{equation}
-\Lambda(3\gamma-2)\left[a^{\gamma}b^{2(\gamma-1)}
\right]^{-(4\gamma-2)/(3\gamma-2)}\int(ab^{2})^{\gamma}d[a^{\gamma}
b^{2(\gamma-1)}]^{2\gamma/(3\gamma-2)}
\end{equation}

when $\gamma>1$. The 2nd, 3rd  and  4th  terms  are negligible
compare to the first term, hence  we  have

\begin{equation}
\frac{\Omega}{\Omega-1}\sim\frac{2\gamma}{5\gamma-2}\left[a^{\gamma}b^{2(\gamma-1)}
\right]^{(4\gamma-3)/(3\gamma-2)}
\end{equation}

So, for large  $a, b$ (i.e., $a, b\rightarrow\infty$)

$$
\frac{\Omega}{\Omega-1}\rightarrow\infty~~~~i.e.,~~\Omega\rightarrow
1.
$$

Furthermore, if $\Lambda=0$ then

\begin{equation}
\frac{\Omega}{\Omega-1}=\frac{2\gamma}{5\gamma-2}\left[a^{\gamma}b^{2(\gamma-1)}
\right]^{(4\gamma-3)/(3\gamma-2)}+\frac{8\pi G
B}{c_{0}^{2}}\left[a^{\gamma}b^{2(\gamma-1)}
\right]^{-(4\gamma-2)/(3\gamma-2)}
\end{equation}

So for the radiation era  ($\gamma=4/3$)

\begin{equation}
\frac{\Omega}{\Omega-1}=\frac{8}{14}+\frac{8\pi G
B}{c_{0}^{2}}(a^{2}b)^{-14/9}.
\end{equation}

As $a, b\rightarrow\infty,~\Omega/(\Omega-1)\rightarrow 8/14$
i.e., $\Omega=-4/3$. This  asymptotic  behaviour  is  same  as
in  Sec.VI$A$ where  for quasi-flatness  problem  the  BD  scalar
field  is  assumed  to be  constant.

\section{\normalsize\bf{Discussion}}

In this work an extensive analysis of the BD solutions for
anisotropic cosmological model has been done where there is a
variation of the velocity of light with time we have assumed the
velocity of light to be in power-law form in the sacle factors.
The flatness problem has been discussed in details. Here
perturbative, non-perturbative and exact solutions of flatness
problem have been obtained. The graphical representation for
non-perturbative solution in the radiation era has some
interesting feature. Quasi-flatness problem with general
asymptotic bahaviour has also been discussed and in most cases we
obtain an open universe. Finally, we have also studied the lambda
and quasi-lambda problem in VSL and the asymptotic behaviour is
very similar to that in the flatness problem.\\

{\bf Acknowledgement:}\\

The authors are thankful to the Relativity and Cosmology Research
Centre, Department of Physics, Jadavpur University for helpful
discussion. One of the authors (U.D) is thankful to CSIR (Govt.
of India) for providing him with a Junior Research Fellowship.\\

{\bf References:}\\
\\
$[1]$  A. Albrecht and J. Magueijo, {\it Phys. Rev. D} {\bf 59} 043516 (1999).\\
$[2]$  J. D. Barrow and J. Magueijo, {\it Phys. Lett. B} {\bf 443} 104 (1998);
{\it Phys. Lett. B}{\bf 447} 246 (1999).\\
$[3]$  J. D. Barrow and J. Magueijo, {\it Class. Quantum Grav.} {\bf 16} 1435 (1999).\\
$[4]$  J. D. Barrow and J. Magueijo, {\it Astrophys. J. Lett.} {\bf 532} L87 (2000).\\
$[5]$  J. Magueijo, {\it Phys. Rev. D} {\bf 62} 103521 (2000). \\
$[6]$  T. Jacobson and D. Meltingly, {\it gr-qc}/0007031 .\\
$[7]$  L. Landau, P. Sisterna and H. Vecetich, {\it Astro-ph}/0007108.\\
$[8]$  B. Basset, {\it et al., Phys. Rev. D} {\bf 62} 103518 (2000).\\
$[9]$  K. Kiritsis, {\it J. High Energy Phys.} {\bf 10} 010 (1999).\\
$[10]$ T. Harko and M. K. Mak, {\it Gen. Rel. Grav.} {\bf 31} 849 (1999);
{\it Class. Quantum Grav.} {\bf 16} 2741 (1999).\\
$[11]$ J. Magueijo, {\it Phys. Rev. D} {\bf 63} 043502 (2001).\\
$[12]$ J. D. Barrow, {\it Phys. Rev. D} {\bf 59} 043515 (1999).\\

\end{document}